\documentclass[12pt, draftclsnofoot, onecolumn]{IEEEtran}
\usepackage{graphicx}
\usepackage{amsmath}
\usepackage[noend]{algorithmic}
\usepackage{algorithm}
\usepackage{amsfonts}
\usepackage{amssymb}
\usepackage{bm}
\usepackage{cite}
\usepackage{color}
\usepackage{multirow}
\usepackage{tabularx}
\newtheorem{lem}{Lemma}

\newcolumntype{L}[1]{>{\raggedright\arraybackslash}p{#1}}
\newcolumntype{C}[1]{>{\centering\arraybackslash}p{#1}}
\newcolumntype{R}[1]{>{\raggedleft\arraybackslash}p{#1}}
\newtheorem{remark}{Remark}
\usepackage{times, epsfig}
\usepackage{subfig}
\newlength{\figwidth}
\usepackage{booktabs}

\usepackage{geometry}
\begin{document}
\title{
Performance Comparison between Reconfigurable Intelligent Surface and Relays:
Theoretical Methods and a Perspective from Operator
}

\author{
Qi Gu$^\dag$, Dan Wu$^\dag$, Xin Su$^\dag$, Jing Jin$^\dag$, Yifei Yuan$^\dag$, and Jiangzhou Wang$^\ast$\\
$^\dag$ Future Mobile Technology Laboratory, China Mobile Research Institute, Beijing, China\\
$^\ast$ School of Engineering and Digital Arts, University of Kent, Canterbury, U.K.\\
 E-mail: $^\dag$\{guqi, wudan, suxin, jinjing, yuanyifei\}@chinamobile.com,
$^\ast$j.z.wang@kent.ac.uk
}
\maketitle
\thispagestyle{empty}
\pagestyle{empty}

\begin{abstract}
Reconfigurable intelligent surface (RIS) is an emerging technique employing metasurface to reflect the signal from the source node to the destination node without consuming any energy.
Not only the spectral efficiency but also the energy efficiency can be improved through RIS.
Essentially, RIS can be considered as a passive relay between the source and destination node.
On the other hand, a relay node in a traditional relay network has to be active, which indicates that it will consume energy when it is relaying the signal or information between the source and destination nodes.
In this paper, we compare the performances between RIS and active relay for a general multiple-input multiple-output (MIMO) system.
To make the comparison fair and comprehensive, both the performances of RIS and active relay are optimized with best-effort.
In terms of the RIS, transmit beamforming and reflecting coefficient at the RIS are jointly optimized so as to maximize the end-to-end throughput.
Although the optimization problem is non-convex, it is transformed equivalently to a weighted mean-square error (MSE) minimization problem and an alternating optimization problem is proposed, which can ensure the convergence to a stationary point.
In terms of active relay, both half duplex relay (HDR) and full duplex relay (FDR) are considered. End-to-end throughput is maximized via an alternating optimization method.
Numerical results are presented to demonstrate the effectiveness of the proposed algorithm.
Finally, comparisons between RIS and relays are investigated from the perspective of system model, performance, deployment and controlling method.

\end{abstract}


\section{Introduction} \label{s:intro}

From the deployment experiences of long term evolution (LTE) relay in operator's network, the most powerful competitor for relay is the repeaters, which can be viewed as an amplify-and-forward relay. Compared with relay specified in 3GPP Release 10 \cite{wg2010evolved}, the repeaters have the advantage of low cost, despite that the noise floor would be raised by the repeaters.
Another advantage is that deploying repeaters in the network does not require significant software or hardware update for core network and base stations, which has relatively low implementation complexity.
One of the most important application scenarios for relays is to provide wireless backhaul when there is no fiber connection or hard to deploy fibers.
However, based on the fiber coverage data in China published by the Ministry of Industry and Information Technology (MIIT) of China, by the year 2019, the fiber and LTE coverage has already reached 98\% across administrative villages throughout China.
This inevitably reduces the demand for relay deployment.
Another issue is that relay needs power supply, and it is often difficult to provide electric power in places where optical fiber is difficult to deploy.
In other application scenarios, for example, in hot spot areas such as stadiums or exhibition halls, where relays are expected for throughput enhancement, the difficulty of reaching power supply is not less than that of optical fiber connection.
If the power is available, it would be easier for operators to deploy a node in the form of a small cell or Remote Radio Unit (RRU) than a relay to achieve both coverage and capacity enhancement, without making substantial changes to the network.
As a result, in addition to performance, power supply is one of the most important factors that limits the development of technologies such as relay or repeaters.
On the other hand, RIS is an emerging technique employing metasurface to reflect the signal from the source node to the destination node without consuming any energy \cite{wu2019towards,zhang2020capacity,wu2019intelligent}.
Not only the spectral efficiency but also the energy efficiency can be improved through RIS, which can be regarded as a technology that can be immune to power supply limitation.
Considering the low cost and low power features \cite{subrt2012intelligent}, it may be possible to install passive or semi-passive control of RIS that is free from the wired connections.

However, the research on comparison of RIS and relays is at state of early age \cite{bjornson2019intelligent,boulogeorgos2020performance,ntontin2020rate}.
In \cite{bjornson2019intelligent}, RIS-assisted single input single output (SISO) systems was compared with classic decode and forward (DF) relaying, which ignored the small scale fading.
RIS-assisted  SISO systems was compared with amplify and forward (AF) relaying wireless systems in \cite{boulogeorgos2020performance}.
\cite{ntontin2020rate} considered the insertion losses and power consumption of
the electronic components related with the deployed nodes for  energy efficiency.

In this paper, we compare the RIS and its counterpart FDR/HDR from the aspects of system models, performance and control aspects.
The basic question we would like to answer is what the fundamental difference between RIS and relay from both theoretical  industrial views.
We first analyze the system models and  end-to-end throughput maximization problem for RIS-aided, FDR-aided and HDR-aided multiple input multiple output (MIMO) systems.
Then an alternating weighted minimum mean square error (MMSE) algorithm is proposed for finding approximate solutions.
Some simulation results are provided to demonstrate its efficacy.
Finally, we discuss the comparisons of RIS and relays.

{\it Notation:}
$\mathbb{C}^{N \times M }$ represents the set of complex ${N \times M}$-matrix.
The conjugate transpose is denoted by $(\cdot)^H$.
$ \mathcal{CN}$(${\bm\mu}$, $\mathbf{\Sigma}$) denotes the circularly symmetric complex Gaussian distribution with mean ${\bm\mu}$ and covariance matrix $\mathbf{\Sigma}$.
$\mathrm{tr}(\cdot)$   represents the trace of matrix.
$\mathbf{I}_N $ denotes the $N\times N$ identity matrix.
$\mathbf{X} \succeq \mathbf{0}$  means that $\mathbf{X}$ is a positive semidefinite matrix.

\section{System Model and Problem Formulation} \label{s:model}
In this section, we describe the system models and formulate the optimization problems for MIMO communication system assisted by RIS, full duplex relay and half duplex relay, respectively.

\subsection{Reconfigurable Intelligent Surface}
\begin{figure}[!hbtp]
\begin{center}
\includegraphics[angle=0,width=0.5\textwidth]{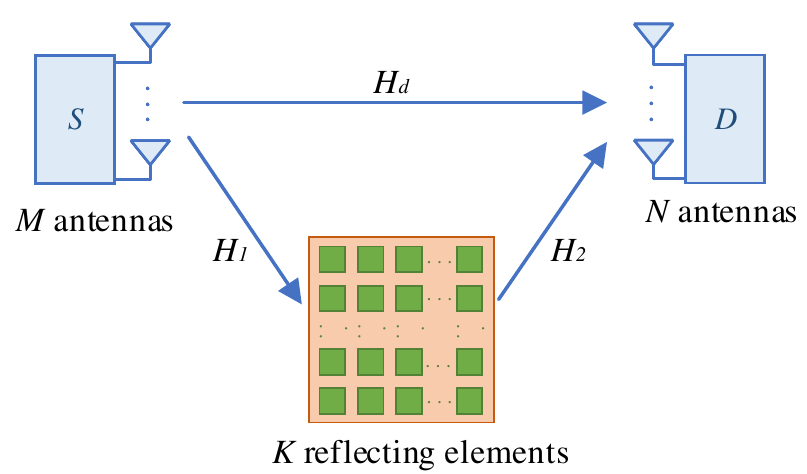}
\end{center}
\caption{An illustration of a RIS-aided MIMO communication system. }
\label{fig:system_RIS}
\end{figure}

Consider a RIS-aided MIMO communication system where the transmitter is equipped with $M$ antennas and the receiver is equipped with $N$ antennas as shown in Fig. \ref{fig:system_RIS}. The RIS is equipped with $K$ reflecting elements.
The channel between the transmitter to the receiver is denoted as $\mathbf{H}_d \in \mathbb{C}^{N \times M} $.
Denote $\mathbf{H}_1 \in \mathbb{C}^{K \times M} $ as the channel matrix from the transmitter to the RIS, $\mathbf{H}_2 \in \mathbb{C}^{N \times K} $ as the channel matrix from the RIS to the receiver.

The transmitted signal ${\bm x} \in \mathbb{C}^{M \times 1}$ at the transmitter can be represented by
\begin{align} \label{eq:xvs}
{\bm x}= \mathbf{V}  {\bm s},
\end{align}
where $\mathbf{V} \in \mathbb{C}^{M \times l} $ is the transmit beamforming matrix, ${\bm s} \in \mathbb{C}^{l \times 1}$ is the transmit data,
$l$ is the number of data streams, assuming $\mathbb{E}\{{\bm s}{\bm s}^H\}=\mathbf{I}$.
Consider the power constraint $\mathrm{tr}(\mathbf{V}\mathbf{V}^H)\leq P_s$, where $P_s$  is the power budget of the transmitter.
The received signal ${\bm y} \in \mathbb{C}^{N \times 1}$ at the receiver is given by
\begin{align}
{\bm y}= (\mathbf{H}_d +\mathbf{H}_2 \mathbf{\Phi} \mathbf{H}_1) \mathbf{V}  {\bm s} + {\bm n}_1,
\label{eq:y_ris}
\end{align}
where
$ {\bm n}_1 \sim \mathcal{CN}(0, \sigma_D^2 \mathbf{I}_K) $ is an additive white Gaussian noise (AWGN) with zero mean and variance $\sigma_D^2 \mathbf{I}_N$,
the diagonal matrix $\mathbf{\Phi} = \text{diag} (\phi_1, \phi_2,...,\phi_k)$ denotes RIS reflection coefficients.
The estimated signal $\hat{\bm s}$ at receiver is given by
\begin{align}
\hat{\bm s} = \mathbf{U}  {\bm y},
\end{align}
where $\mathbf{U} $ is the receive beamforming matrix.
The throughput maximization of the RIS-aided MIMO communication system can be given by
\begin{subequations} \label{P1}
\begin{align}
\max \limits_{\mathbf{V},\mathbf{\Phi}}~ & \log \det \left(  \mathbf{I}_N + \mathbf{H}\mathbf{V}\mathbf{V}^H \mathbf{H}^H \mathbf{R}_n^{-1} \right)\\
\text{s.t.} \quad
&  \mathrm{tr}(\mathbf{V}\mathbf{V}^H)\leq P_t,\\
&  | \phi_k|\leq 1,
\end{align}
\end{subequations}
where $\mathbf{H}=\mathbf{H}_d +\mathbf{H}_2 \mathbf{\Phi} \mathbf{H}_1$, $\mathbf{R}_n^{-1} = \sigma_D^2 \mathbf{I}_K$.
It can be seen that the objective function of problem \eqref{P1} involves the multiplication of $\mathbf{\Phi}$ and $\mathbf{V}$,
which is a nonconvex function. Thus problem \eqref{P1} is a non-convex optimization problem, the global optimal solution of which is hard to calculate.

\subsection{Full Duplex Relay}
\begin{figure}[!hbtp]
\begin{center}
\includegraphics[angle=0,width=0.5\textwidth]{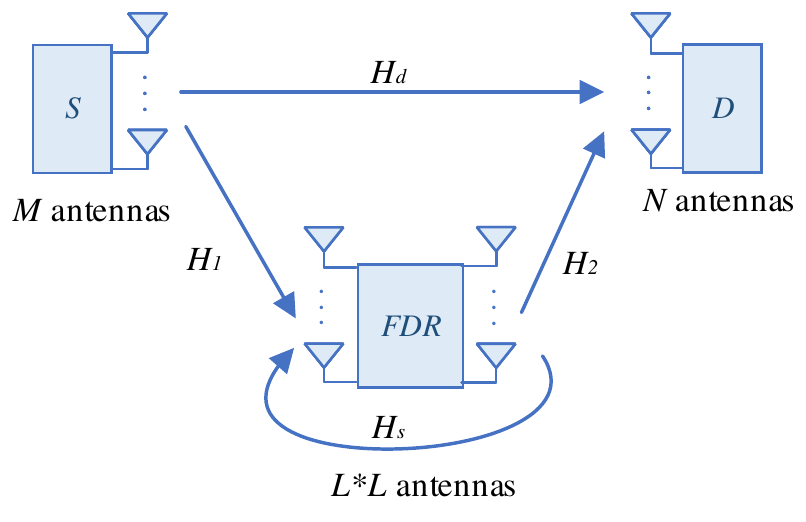}
\end{center}
\caption{An illustration of a FDR-aided MIMO communication system. }
\label{fig:system_FDR}
\end{figure}

As shown in Fig. \ref{fig:system_FDR}, we consider a MIMO communication system assisted by a FDR that suffers from self-interference, where $\mathbf{H}_s \in \mathbb{C}^{L \times L}$ is the channel matrix.
The transmitted signal ${\bm x} $ at the transmitter is the same as \eqref{eq:xvs}.
The transmitted signal ${\bm x}_r$ at the FDR is given by
\begin{align} \label{eq_x_r}
{\bm x}_r = \mathbf{G} (\mathbf{H}_1 {\bm x}+\mathbf{H_s} {\bm x}_r + {\bm n}_1),
\end{align}
where $\mathbf{G} \in \mathbb{C}^{L \times L} $  is the FDR transmit matrix.
According to \eqref{eq_x_r}, we can obtain
\begin{align} \label{eq_x_r_2}
{\bm x}_r = (\mathbf{I}_k- \mathbf{G}\mathbf{H}_s)^{-1}\mathbf{G}(\mathbf{H}_1 {\bm x}+{\bm n_1}).
\end{align}
Defining $\mathbf{F}=(\mathbf{I}_R-\mathbf{G}\mathbf{H}_s)^{-1}\mathbf{G}$,
the received signal ${\bm y}$ at the receiver is given by
\begin{align}
{\bm y}&=(\mathbf{H}_d+\mathbf{H}_2\mathbf{F}\mathbf{H}_1)\mathbf{V} { \bm s}+ \mathbf{H}_2\mathbf{F} {\bm n}_1+ {\bm n}_2.
\end{align}
The throughput maximization problem of the FDR-aided MIMO communication system is given by
\begin{subequations} \label{P_FDR}
\begin{align}
\max \limits_{\mathbf{V},\mathbf{F}}~ & \log\det\left(\mathbf{I}_N+ \mathbf{H}\mathbf{V}\mathbf{V}^H
\mathbf{H}^H\mathbf{R}_n^{-1}\right) \\
\text{s.t.} \quad
&  \text{tr}(\mathbf{V}\mathbf{V}^H)\leq P_s,\\
&  \text{tr}(\mathbf{F} \mathbf{D} \mathbf{F}^H)
\leq P_r,
\end{align}
\end{subequations}
where $\mathbf{R}_n=\sigma_R^2\mathbf{H}_2\mathbf{F}\mathbf{F}^H\mathbf{H}_2^H +\sigma_D^2\mathbf{I}_N$,
define
$ \mathbf{H}=\mathbf{H}_d+ \mathbf{H}_2\mathbf{F}\mathbf{H}_1 $,
$\mathbf{D}=(\mathbf{H}_1\mathbf{V}\mathbf{V}^H\mathbf{H}_1^H +\sigma_R^2\mathbf{I})$
 for clarity.

\subsection{Half Duplex Relay}
\begin{figure}[!hbtp]
\begin{center}
\includegraphics[angle=0,width=0.5\textwidth]{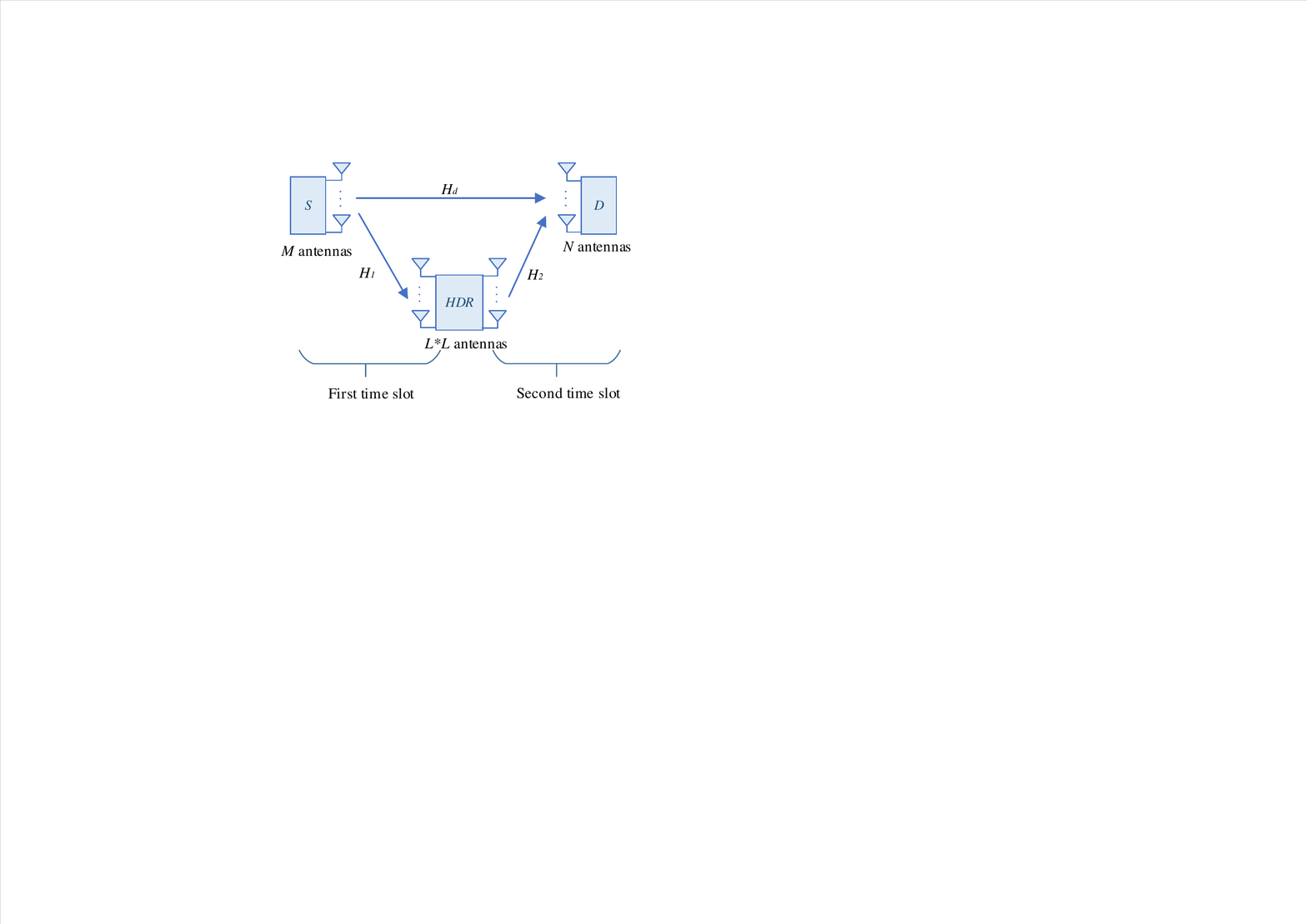}
\end{center}
\caption{Illustration of a HDR-aided MIMO communication system. }
\label{fig:system_HDR}
\end{figure}

We consider the HDR aided MIMO communication system as comparison as shown in Fig. \ref{fig:system_HDR}.
The received signal ${\bm y}$ is given by
\begin{align}
{\bm y}&=(\mathbf{H}_d+\mathbf{H}_2\mathbf{G}\mathbf{H}_1)\mathbf{V} { \bm s}+ \mathbf{H}_2\mathbf{G} {\bm n}_1+ {\bm n}_2,
\end{align}
where $\mathbf{G} \in \mathbb{C}^{L \times L} $ is the HDR transmit matrix.
For clarity, defining
$ \mathbf{H}=\mathbf{H}_d+ \mathbf{H}_2\mathbf{G}\mathbf{H}_1 $,
the corresponding throughput maximization problem is given by
\begin{subequations} \label{P_HDR}
\begin{align}
\max \limits_{\mathbf{V},\mathbf{G}}~ &\frac{1}{2}\log\det\left(\mathbf{I}_N+ \mathbf{H}\mathbf{V}\mathbf{V}^H\mathbf{H}^H\mathbf{R}_n^{-1}\right)\\
\text{s.t.} \quad
& \text{tr}(\mathbf{G}(\mathbf{H}_1 \mathbf{V}\mathbf{V}^H \mathbf{H}_1^H+\sigma_R^2\mathbf{I})\mathbf{G}^H)\leq2P_r, \\
&\text{tr}(\mathbf{V}\mathbf{V}^H)\leq2P_s,
\end{align}
\end{subequations}
where $\mathbf{R}_n=\sigma_R^2\mathbf{H}_2\mathbf{G}\mathbf{G}^H\mathbf{H}_2^H +\sigma_D^2\mathbf{I}_N$.
Since the transmission needs two time slots for one transmission, the objective function and constraints have the factor $\frac{1}{2}$ and $2$, respectively \cite{kang2009capacity}.

\begin{remark}\label{remark_1}
From the system models, it can be observed that RIS can be regarded as a full-duplex MIMO relay without self-interference.
\end{remark}

\section{Alternating Weighted MMSE Approach }
\subsection{Reconfigurable Intelligent Surface}
Assuming the signal ${\bm s}$ is independent of the noise ${\bm n}_1$ in \eqref{eq:y_ris},
the MSE covariance matrix $\mathbf{E}$ can be represented by
\begin{align}
\mathbf{E}&=\mathbb{E}\{(\hat{{\bm s}}-{\bm s})(\hat{{\bm s}}-{\bm s})^H\} \nonumber \\
&=
(\mathbf{U}^H\mathbf{H}\mathbf{V-\mathbf{I}})
(\mathbf{U}^H\mathbf{H}\mathbf{V}-\mathbf{I})^H+\sigma_D^2 \mathbf{U}^H\mathbf{U}.
\end{align}
The MSE minimization problem is to solve
\begin{subequations} \label{P_MMSE}
\begin{align}
\min \limits_{\mathbf{V},\mathbf{\Phi},\mathbf{U}}~ & \mathrm{tr}(\mathbf{E})  \\
\text{s.t.} \quad
&  \mathrm{tr}(\mathbf{V}\mathbf{V}^H)\leq P_t,\\
&  | \phi_k|\leq 1.
\end{align}
\end{subequations}
Assuming that the transmit beamforming $\mathbf{V}$ and RIS reflection coefficients $\mathbf{\Phi}$ are fixed,
the receive beamforming $\mathbf{U}$ in the MMSE receiver is written by
\begin{align} \label{eq_U_mmse}
\mathbf{U}_{\text{mmse}}=(\mathbf{H}\mathbf{V}\mathbf{V}^H\mathbf{H}^H +\sigma_D^2\mathbf{I})^{-1}\mathbf{H}\mathbf{V}.
\end{align}
The corresponding MSE matrix $\mathbf{E}$ can be represented by
\begin{align} \label{eq_E_mmse}
\mathbf{E}_{\text{mmse}}&=\mathbf{I}-\mathbf{V}^H\mathbf{H}^H \mathbf{R}_n^{-1}
(\mathbf{H}\mathbf{V}\mathbf{V}^H\mathbf{H}^H\mathbf{R}_n^{-1} +\mathbf{I})^{-1}\mathbf{HV}
\end{align}

The weighted MSE minimization problem is given by
\begin{subequations} \label{P_WMMSE}
\begin{align}
\min \limits_{\mathbf{V},\mathbf{\Phi},\mathbf{U},\mathbf{W}}~ & \mathrm{tr}(\mathbf{WE}) - \log \det (\mathbf{W})  \label{P_WMMSE_obj}\\
\text{s.t.} \quad
&  \mathrm{tr}(\mathbf{V}\mathbf{V}^H)\leq P_s,\\
&  | \phi_k|\leq 1,
\end{align}
\end{subequations}
where $\mathbf{W}\succeq 0$ is a weight matrix.

\begin{lem} \label{lemma1}
The Weighted MSE minimization problem \eqref{P_WMMSE} is shown to
be equivalent to the throughput maximization problem \eqref{P1}.
\end{lem}
\begin{IEEEproof}
First, the optimal $\mathbf{U}$ and $\mathbf{E}$ of problem \eqref{P_WMMSE} are given in \eqref{eq_U_mmse} and \eqref{eq_E_mmse}, respectively.
When other variables are given, the objective function of problem \eqref{P_WMMSE} is convex with respect to (w.r.t) $\mathbf{W}$.
Therefore, we can obtain
$\mathbf{W}= \mathbf{E}_{\text{mmse}}^{-1}$ by checking first order optimality condition.
Substituting $\eqref{eq_U_mmse}$ and $\mathbf{W}$ into \eqref{P_WMMSE_obj},
we have
\begin{align}
\min \limits_{\mathbf{V},\mathbf{\Phi}}~ & - \log \det (\mathbf{E}_{\text{mmse}}^{-1})
\nonumber \\
=\max \limits_{\mathbf{V},\mathbf{\Phi}}~ &  \log \det (\mathbf{E}_{\text{mmse}}^{-1})
\nonumber \\
\overset{\text{(a)}}{=} \max \limits_{\mathbf{V},\mathbf{\Phi}} ~ &
\log \det (\mathbf{I}+\mathbf{V}^H\mathbf{H}^H\mathbf{R}_n^{-1}\mathbf{HV})^{-1}, \nonumber \\
\overset{\text{(b)}}{=} \max \limits_{\mathbf{V},\mathbf{\Phi}}~ & \log \det \left(  \mathbf{I}_N + \mathbf{H}\mathbf{V}\mathbf{V}^H \mathbf{H}^H \mathbf{R}_n^{-1} \right),
\end{align}
where (a) comes from equality $\mathbf{I}-\mathbf{B}(\mathbf{AB}+\mathbf{I})^{-1}\mathbf{A}=(\mathbf{I}+\mathbf{BA})^{-1}$,
(b) comes from equality  $\det(\mathbf{I+\mathbf{AB}})=\det(\mathbf{I+\mathbf{BA}})$.
\end{IEEEproof}

Since problem \eqref{P1} is nonconvex, it is difficult to handle.
Therefore, we can handle the problem \eqref{P_WMMSE} by utilizing  alternating optimization.
The objective function of problem \eqref{P_WMMSE} is convex  w.r.t  each of the optimization variables
$\{\mathbf{V},\mathbf{\Phi},\mathbf{W},\mathbf{U}\}$.

When $\mathbf{V},\mathbf{\Phi}$ are given, the optimal $\mathbf{U}$ can be obtained from \eqref{eq_U_mmse}.
With \eqref{eq_E_mmse}, the weight matrix $\mathbf{W}$ is given by
\begin{align} \label{eq_W_mmse}
\mathbf{W}= \mathbf{E}_{\text{mmse}}^{-1}.
\end{align}

When $\mathbf{\Phi},\mathbf{U},\mathbf{W}$ are given, by substituting \eqref{eq_U_mmse} and \eqref{eq_E_mmse},
problem \eqref{P_WMMSE} can be reformulated as
\begin{subequations} \label{P_WMMSE_UWPhi}
\begin{align}
\min_{\mathbf{V}}~ & \text{tr}(\mathbf{V}^H\mathbf{H}^H\mathbf{U}\mathbf{W} \mathbf{U}^H\mathbf{H}\mathbf{V})
 -\text{tr}(\mathbf{W}\mathbf{U}^H\mathbf{H}\mathbf{V}) \nonumber \\
&  \quad  -\text{tr}(\mathbf{W}\mathbf{V}^H\mathbf{H}^H\mathbf{U})  \\
\text{s.t.} ~&\text{tr}(\mathbf{V}\mathbf{V}^H)\leq P_s,
\end{align}
\end{subequations}
which is a convex problem and can be efficiently solved by off-the-shelf convex solvers (e.g. CVX and SeDuMi) \cite{boyd2004convex}.

When $\mathbf{U},\mathbf{W},\mathbf{V}$ are given,
the objective function  can be reformulated as follows
\begin{align}\label{eq_f_phi}
f(\mathbf{\Phi})=
\text{tr} (\mathbf{W}
(\mathbf{U}^H\mathbf{H}_d\mathbf{V}\mathbf{V}^H\mathbf{H}_1^H \mathbf{\Phi}^H\mathbf{H}_2^H\mathbf{U}
+\mathbf{U}^H\mathbf{H}_2\mathbf{\Phi}\mathbf{H}_1\mathbf{V} \mathbf{V}^H\mathbf{H}_d^{H}\mathbf{U}
+\mathbf{U}^{H}\mathbf{H}_2\mathbf{\Phi}\mathbf{H}_1\mathbf{V} \mathbf{V}^H\mathbf{H}_1^{H}\mathbf{\Phi}^H\mathbf{H}_2^H\mathbf{U})
\nonumber\\
-\mathbf{W}(\mathbf{V}^H\mathbf{H}_1^H\mathbf{\Phi}^H \mathbf{H}_2^H\mathbf{U}+
\mathbf{U}^H\mathbf{H}_2\mathbf{\Phi}\mathbf{H}_1\mathbf{V}))
\end{align}
Denote $\mathbf{a} = \text{diag} (\mathbf{A})$,
then there is
\begin{align} \label{eq:20210121_0045}
\text{tr}(\mathbf{A}^H \mathbf{B} \mathbf{A} \mathbf{C} )=\mathbf{a} ^H( \mathbf{C} \odot \mathbf{B}^T )\mathbf{a} ,
\end{align}
where $\odot$ is the operation of taking the real part.
With the aid of equality \eqref{eq:20210121_0045},
problem \eqref{P_WMMSE} then becomes
\begin{subequations} \label{P_WMMSE_UWV}
\begin{align}
\min_{{\bm \phi}}~&  \ \ {\bm \phi}^H\mathbf{\Xi}{\bm \phi}+2\mathcal{R}\{{\bm \phi}^H\mathbf{b} ^*\} \\
\text{s.t.} & \ \ |{\bm \phi}_k|\leq 1, \\
\text{where}& \nonumber \\
\mathbf{\Xi} &=(\mathbf{H}_2^H\mathbf{U}\mathbf{W}\mathbf{U}^H\mathbf{H}_2)\odot
(\mathbf{H}_1\mathbf{V}\mathbf{V}^H\mathbf{H}_1^H)^T, \\
\mathbf{B}
&=\mathbf{H}_1\mathbf{V}(\mathbf{V}^H\mathbf{H}_d^H\mathbf{U}
-\mathbf{I})\mathbf{W}\mathbf{U}^H\mathbf{H}_2,\\
\mathbf{b} &=\text{diag}(\mathbf{B}),
\end{align}
\end{subequations}
$\mathcal{R}$ is the operation of taking the real part.
Problem \eqref{P_WMMSE_UWV} is also a convex problem.

Alternating optimization algorithm for obtaining an approximate solution to problem \eqref{P_WMMSE} is summarized in Algorithm \ref{alg1} below.

\begin{algorithm}[H]
\caption{Alternating optimization algorithm}
\label{alg1}
\begin{algorithmic}[1]
 \STATE
  Initialize a feasible $\{\mathbf{V}^1,\mathbf{\Phi}^1 \} $  to  problem \eqref{P_WMMSE};
 \REPEAT
 \STATE
  For given $\{\mathbf{V}^{i},\mathbf{\Phi}^{i} \} $ , the optimal $\{\mathbf{U}^{i},\mathbf{W}^{i} \} $ are given by \eqref{eq_U_mmse} and  \eqref{eq_W_mmse}, respectively.
 \STATE
 Solve Probelm  \eqref{P_WMMSE_UWPhi} for given $\{\mathbf{\Phi}^{i},\mathbf{U}^{i}, \mathbf{W}^{i} \}$ by convex optimization, denote the solution as $\mathbf{V}^{i+1}$.
 \STATE
 Solve Probelm  \eqref{P_WMMSE_UWV} for given $\{\mathbf{U}^{i}, \mathbf{W}^{i},\mathbf{V}^{i+1} \}$ by convex optimization, denote the solution as $\mathbf{\Phi}^{i+1} = \text{diag} ({\bm \phi})$.
 \STATE
  set $i:=i+1$;
 \UNTIL $| f(\mathbf{\Phi}^{i})-f(\mathbf{\Phi}^{i-1})| \leq  \epsilon$.
 \STATE
$\{\mathbf{V}^{i},\mathbf{\Phi}^{i},\mathbf{U}^{i-1},\mathbf{W}^{i-1}\}$ is  the obtained approximate  solution of problem \eqref{P_WMMSE}.
\end{algorithmic}
\end{algorithm}

\begin{remark}\label{remark_2}
The optimal solution generated by alternating optimization algorithm 1 is a stationary point of weighted MSE minimization problem \eqref{P_WMMSE}.
The proof is omitted due to limited space.
Similar proof can be found in \cite{shi2011iteratively}.
\end{remark}

\subsection{Full Duplex Relay}
Similarly, the throughput maximization problem \eqref{P_FDR} is equivalent to the weighted MSE minimization problem as follows
\begin{subequations} \label{P_WMMSE_FDR}
\begin{align}
\min \limits_{\mathbf{V},\mathbf{F},\mathbf{U},\mathbf{W}}~ & \mathrm{tr}(\mathbf{WE}) - \log \det (\mathbf{W})  \\
\text{s.t.} \quad
&  \text{tr}(\mathbf{V}\mathbf{V}^H)\leq P_s,\\
& \text{tr}(\mathbf{F}\mathbf{D}\mathbf{F}^H)\leq P_r.
\end{align}
\end{subequations}

When $\mathbf{V},\mathbf{F}$ are given, the optimal $\mathbf{U}$, $\mathbf{W}$  can be obtained as
\begin{align}
\mathbf{U}_{\text{mmse}}&=
\left(\mathbf{H}\mathbf{V}\mathbf{V}^H\mathbf{H}^{\mathrm{H}}+
\mathbf{R}_n
\right)^{-1}\mathbf{H}\mathbf{V},\\
\mathbf{W}&=\mathbf{E}^{-1}_{\text{mmse}},
\end{align}
where $\mathbf{E}_{\text{mmse}}=\mathbf{I}-\mathbf{V}^H\mathbf{H}^H \mathbf{R}_n^{-1}
(\mathbf{H}\mathbf{V}\mathbf{V}^H\mathbf{H}^H\mathbf{R}_n^{-1} +\mathbf{I})^{-1}\mathbf{HV}$.

When $\mathbf{F},\mathbf{U},\mathbf{W}$ are given,
the problem \eqref{P_WMMSE_FDR} is convex w.r.t. the transmit beamforming $\mathbf{V}$.

When $\mathbf{U},\mathbf{W},\mathbf{V}$ are given,
the objective function of problem \eqref{P_WMMSE_FDR} can be written as follows
\begin{align}\label{eq_f_F}
g(\mathbf{F})=
\text{tr}(\mathbf{F}^H\mathbf{H}_2^H\mathbf{U}\mathbf{W}\mathbf{U}^H \mathbf{H}_2\mathbf{F}\mathbf{H}_1\mathbf{V}\mathbf{V}^H\mathbf{H}_1^H)+
\text{tr}
(\mathbf{F}^H\mathbf{H}_2^H\mathbf{U}\mathbf{W} (\mathbf{U}^H\mathbf{H}_d\mathbf{V}-\mathbf{I})\mathbf{V}^H \mathbf{H}_1^H)\nonumber\\
+\text{tr}
(\underbrace{\mathbf{H}_1\mathbf{V}(\mathbf{V}^H\mathbf{H}_d^H\mathbf{U} -\mathbf{I})\mathbf{W}\mathbf{U}^H\mathbf{H}_2}_{\mathbf{B}}\mathbf{F})
+\sigma_R^2\text{tr}(\mathbf{F}^H
\underbrace{\mathbf{H}_{2}^H\mathbf{U}\mathbf{W} \mathbf{U}^H\mathbf{H}_2}_{\mathbf{C}}\mathbf{F})
\nonumber \\
=\text{vec}(\mathbf{F})^{\mathrm{H}}\underbrace{ (\mathbf{H}_1\mathbf{V}\mathbf{V}^H\mathbf{H}_1^H)^T \otimes(\mathbf{H}_2^H\mathbf{U}\mathbf{W}\mathbf{U}^H \mathbf{H}_2)}_{\mathbf{A}}
\text{vec}(\mathbf{F})
+2\mathcal{R}\left\{
\underbrace{\text{vec}(\mathbf{B}^H)^H}_{\mathbf{b}^H}\text{vec}(\mathbf{F})
\right\}\nonumber\\
+\sigma_R^2\text{vec}(\mathbf{F})^H(\mathbf{I}_K \otimes\mathbf{C})\text{vec}(\mathbf{F}),
\end{align}
where $\otimes$ is the operation of Kronecker product.
With the aid of equality $\text{tr}(\mathbf{A}^H \mathbf{B} \mathbf{A} \mathbf{C}) = \text{vec}(\mathbf{A})^H \mathbf{C}^T \otimes \mathbf{B}\text{vec}(\mathbf{A})$, the problem \eqref{P_WMMSE_FDR} can be reformulated as
\begin{subequations}  \label{P_WMMSE_FDR_UWV}
\begin{align}
\min_{\mathbf{F}}~ & \mathbf{f}^H\mathbf{\Xi}
\mathbf{f}+ 2\mathcal{R}(\mathbf{b}^H\mathbf{f})\\
\text{s.t.}  \quad
&\mathbf{f}^H(\mathbf{I}\otimes\mathbf{D})\mathbf{f}\leq P_r,
\end{align}
\end{subequations}
in which $\mathbf{\Xi}=\mathbf{A}+\sigma_R^2(\mathbf{I}\otimes\mathbf{C})$,
 $\mathbf{f}=\text{vec}(\mathbf{F})$.
 It can be seen that problem \eqref{P_WMMSE_FDR_UWV} is a convex problem.

Similar to the discussion for problem \eqref{P_WMMSE},
alternating optimization algorithm can be utilized to handle problem \eqref{P_WMMSE_FDR}.

\subsection{Half Duplex Relay}
Problem \eqref{P_HDR} is similar to Problem \eqref{P_FDR}.
Therefore, alternating optimization algorithm 1 can be utilized to solve Problem \eqref{P_HDR} by replacing variable $\mathbf{F}$ with $\mathbf{G}$.

%

\section{Numerical Results and Analysis}
In this section, numerical results are presented to validate the effectiveness of the proposed algorithm.
Besides, the comparisons of RIS and relays are stdudied.

\begin{figure}[h]
\begin{center}
\includegraphics[angle=0,width=0.44\textwidth]{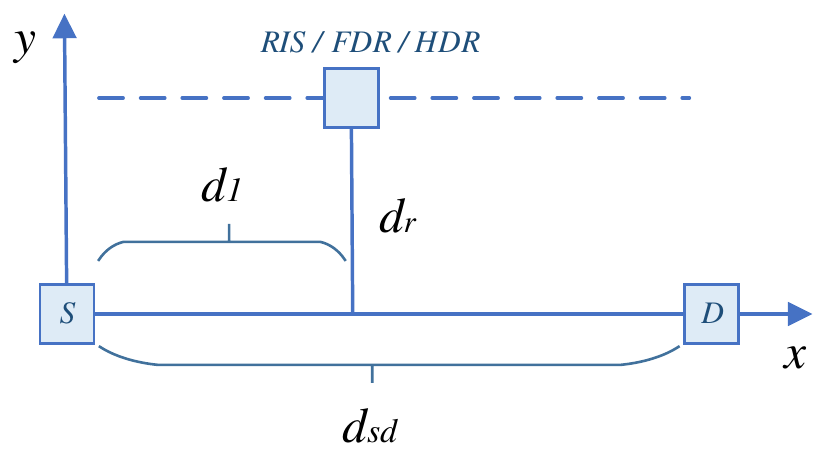}
\end{center}
\caption{The simulation setup. }
\label{fig:setup}
\end{figure}

\subsection{Numerical Results}
The simulation setup is shown in Fig. \ref{fig:setup}.
The location of source node and destination node is set as
(0,0) and ($d_{sd}$,0), respectively.
The location of RIS (or FDR/HDR) is set as $(d_1,d_r)$.
We adopt path loss model from the 3GPP Urban Micro (UMi) \cite{access2010further}
(Table B.1.2.1-1) and Rayleigh fading as small scale fading.
Consider the line-of-sight (LOS) and non-LOS (NLOS) versions
of UMi, which are defined for distances $\geq$ 10 m.
Default system parameters are set as:
$d_{sd}=100$m,
$d_1=50$m,
$d_r=10$m,
$M=4$,
$N=4$,
$L=4$,
$K=200$,
$P_s=P_r=43$dBm,
carrier frequency $f = 3$GHz,
bandwidth $B=100$MHz.

\begin{figure}[!hbtp]
\begin{center}
\includegraphics[angle=0,width=0.56\textwidth]{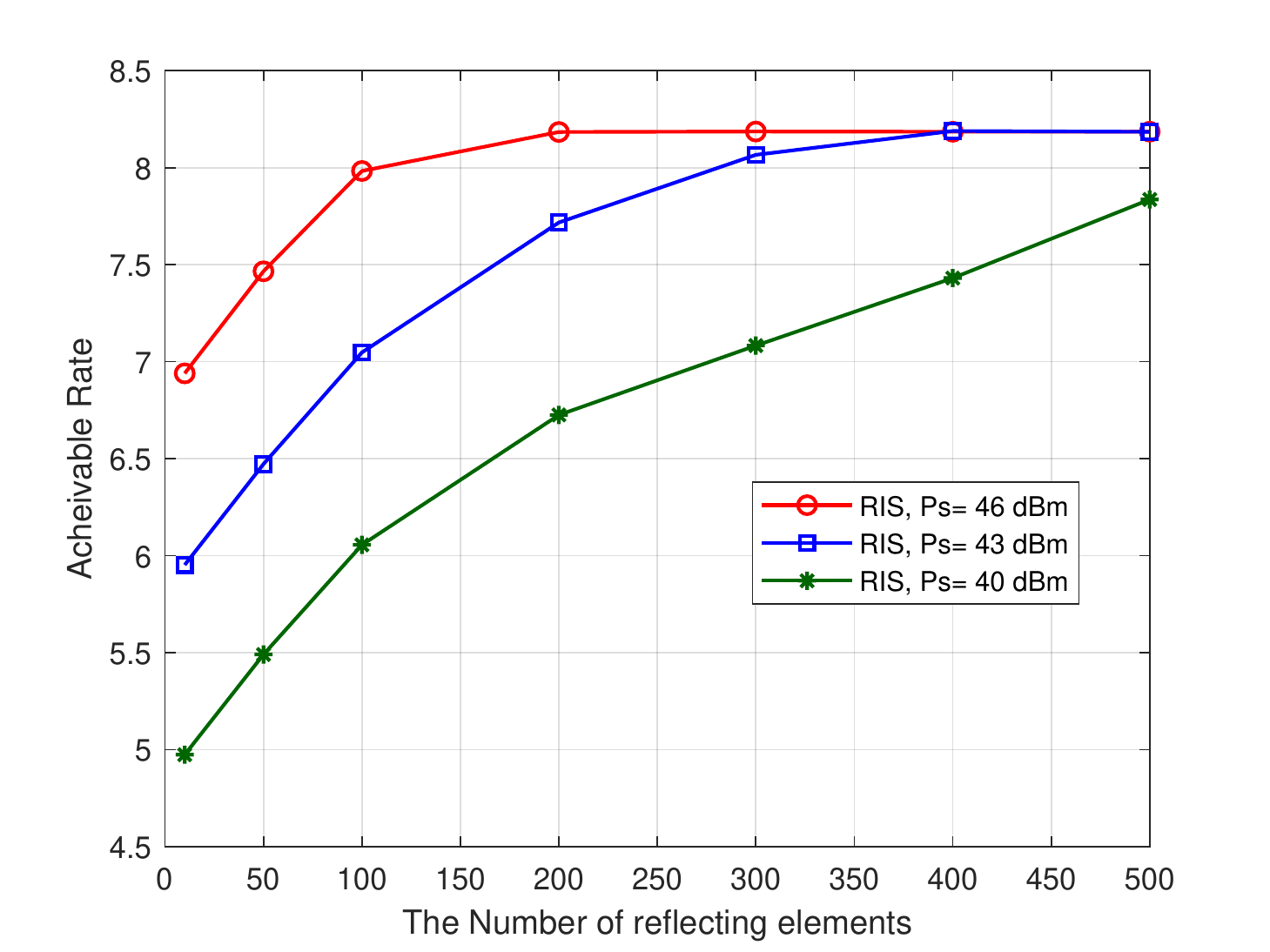}
\end{center}
\caption{Achievable rate versus the number of RIS reflecting elements. }
\label{fig:rate_Num_antenna}
\end{figure}

Fig. \ref{fig:rate_Num_antenna} plots achievable rate versus the number of reflecting elements, i.e. $K$, with different  transmission power  $P_s$.
It can be seen that achievable rate is an increasing function w.r.t $K$.
Besides, RIS performs better for higher $P_s$.

\begin{figure}[!hbtp]
\begin{center}
\includegraphics[angle=0,width=0.5\textwidth]{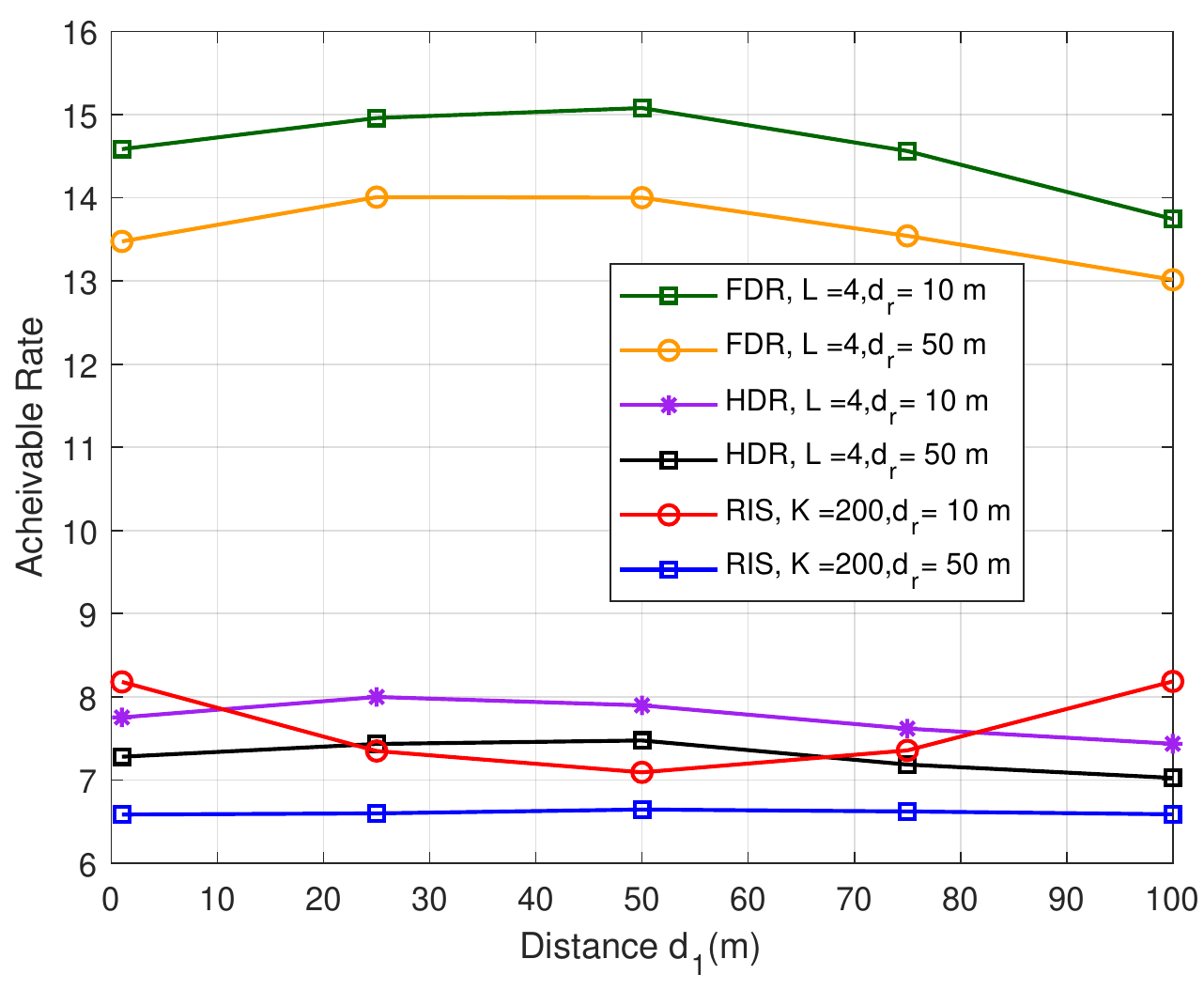}
\end{center}
\caption{Achievable rate versus distance. }
\label{fig:rate_distance1}
\end{figure}
\begin{figure}[!hbtp]
\begin{center}
\includegraphics[angle=0,width=0.5\textwidth]{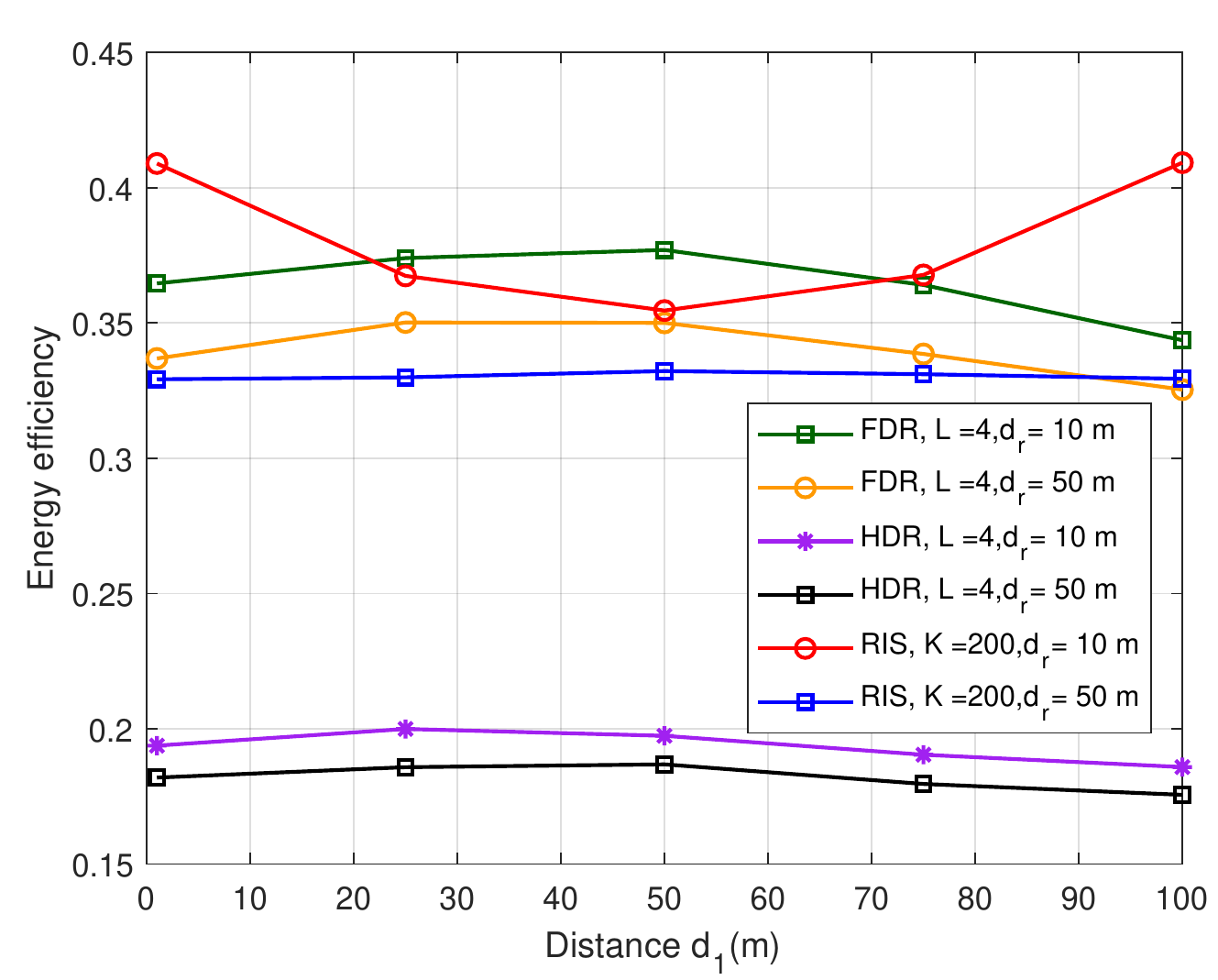}
\end{center}
\caption{Energy efficiency versus distance. }
\label{fig:rate_distance2}
\end{figure}

Fig. \ref{fig:rate_distance1} plots achievable rate versus the distance $d_1$ with different $d_r$ .
Energy efficiency is a very important metric to operator, so we investigate the energy efficiency performance for both RIS and FDR/HDR.
Fig. \ref{fig:rate_distance2} plots energy efficiency versus the distance $d_1$ with different $d_r$.
In Fig. \ref{fig:rate_distance1} and Fig. \ref{fig:rate_distance2}, both transmission power $P_s$ at source node and $P_r$ at relay are set as 43dBm.
From Fig. \ref{fig:rate_distance1} and Fig. \ref{fig:rate_distance2}, we can get the following observations:
\begin{itemize}
\item In terms of achievable rate, FDR performs best, the performance of RIS is neck and neck with the performance of HDR.
\item In terms of energy efficiency,  the performance of RIS is comparable to that of  performance of FDR.
Moreover, the performance of RIS can be better than that of FDR with a large number of reflecting elements and good deployment.
\item Deploying RIS in the vicinity of source node or destination node can achieve better performance.
\end{itemize}

\begin{remark}
According to the above observations, although the RIS does not have a matchable rate compared with FDR, it does not lose FDR in terms of energy efficiency performance, which is an especially important performance metric from the perspective of operator.
Considering the fact that RIS has the advantage in both implementation and energy consumption, this technique is very alluring to the operators.
\end{remark}

\subsection{Comparison between RIS and Relays}
In this section, we discuss the comparison between RIS and relays from different perspectives.
From the perspective of system models, RIS can be regarded as a full-duplex MIMO relay without self-interference.
In addition to the differences in system models, we have also seen differences in perspective of performance with different cases.
Finally, we consider the difference between RIS and relay in actual deployment and application methods from the perspective of operator.
Since RIS can operate in a passive or semi-passive manner, which allows RIS to be deployed in a more flexible way than relay.
As for the loss of power and performance, it can be compensated by increasing the number of reflecting elements. 
The most important feature for RIS is the possibility to be free from power supply.
Only then can it be possible to realize the flexible and on-demand deployment of the sixth generation (6G) network.
It not only is a supplement of coverage, but also provides on-demand service for users and traffic.
To achieve the above goal, the network needs to control the RIS in a wireless way.
New designs are required in terms of protocol architecture and control methods.
Furthermore, the designed solution must ensure low power consumption and low complexity.
There is no restriction on power for relay.
Therefore, the control and processing could be more complicated at relays.

\section{Conclusion and Future Work}
In this paper, we proposed the alternating weighted MMSE algorithm to handle throughput maximization problem by jointly optimizing transmit beamforming and reflecting coefficient of RIS.kang2009capacity
Moreover, the comparisons between RIS and relays are investigated:
\begin{itemize}
\item From the perspective of system models, RIS can be regarded as a full-duplex MIMO relay without self-interference.
\item From the perspective of performance,  RIS is comparable to the spectral efficiency performance of HDR and achieve the same and even better energy efficiency performance than FDR.
\item From the perspective of deployment requirement and controlling method, RIS can provide a low-cost and flexible solution which is free from wired power supply.
\end{itemize}
Our future work includes extension to multiple users MIMO, system level evaluation of RIS-aided network, and system architecture design to achieve low power consumption and low complexity control of RIS.

\end{document}